\documentclass[aps,twocolumn,superscriptaddress,showpacs]{revtex4-1}

\usepackage{graphicx}% Include figure files
\usepackage{dcolumn}% Align table columns on decimal point
\usepackage{bm}% bold math
\usepackage{amsmath}
\usepackage{color}
\usepackage{ulem}
\usepackage{threeparttable}
\begin{document}

\title{Magnetism study on a frustration-free spatially anisotropic $S$ = 1 square lattice antiferromagnet Ni[SC(NH$_2$)$_2$]$_6$Br$_2$}

\author{X. Zhao}
\affiliation{School of Physical Sciences, University of Science and Technology of China, Hefei, Anhui 230026, People's Republic of China}

\author{Z. Y. Zhao}
\email{zhiyingzhao@fjirsm.ac.cn}
\affiliation{State Key Laboratory of Structural Chemistry, Fujian Institute of Research on the Structure of Matter, Chinese Academy of Sciences, Fuzhou, Fujian 350002, People's Republic of China}

\author{L. M. Chen}
\affiliation{College of Electronic Science Engineering, Nanjing University of Post and Telecommunication, Nanjing, Jiangsu 211106, People's Republic of China}

\author{X. Rao}
\affiliation{Department of Physics, Hefei National Laboratory for Physical Sciences at Microscale, and Key Laboratory of Strongly-Coupled Quantum Matter Physics (CAS), University of Science and Technology of China, Hefei, Anhui 230026, People's Republic of China}

\author{H. L. Che}
\affiliation{Department of Physics, Hefei National Laboratory for Physical Sciences at Microscale, and Key Laboratory of Strongly-Coupled Quantum Matter Physics (CAS), University of Science and Technology of China, Hefei, Anhui 230026, People's Republic of China}

\author{L. G. Chu}
\affiliation{Department of Physics, Hefei National Laboratory for Physical Sciences at Microscale, and Key Laboratory of Strongly-Coupled Quantum Matter Physics (CAS), University of Science and Technology of China, Hefei, Anhui 230026, People's Republic of China}

\author{H. D. Zhou}
\affiliation{Department of Physics and Astronomy, University of Tennessee, Knoxville, Tennessee 37996-1200, USA}

\affiliation{Key laboratory of Artificial Structures and Quantum Control (Ministry of Education), School of Physics and Astronomy, Shanghai JiaoTong University, Shanghai, 200240, People's Republic of China}

\author{L. S. Ling}
\affiliation{High Magnetic Field Laboratory, Chinese Academy of Science, Hefei, Anhui 230031, People's Republic of China}

\author{J. F. Wang}
\affiliation{School of physics, Huazhong University of Science and Technology, Wuhan, Hubei 430074, People's Republic of China}

\affiliation{Wuhan National High Magnetic Field Center, Huazhong University of Science and Technology, Wuhan, Hubei 430074, People's Republic of China}

\author{X. F. Sun}
\email{xfsun@ustc.edu.cn}
\affiliation{Department of Physics, Hefei National Laboratory for Physical Sciences at Microscale, and Key Laboratory of Strongly-Coupled Quantum Matter Physics (CAS), University of Science and Technology of China, Hefei, Anhui 230026, People's Republic of China}

\affiliation{Institute of Physical Science and Information Technology, Anhui University, Hefei, Anhui 230601, People's Republic of China}

\affiliation{Collaborative Innovation Center of Advanced Microstructures, Nanjing University, Nanjing, Jiangsu 210093, People's Republic of China}

\date{\today}

\begin{abstract}

Magnetism of the $S$ = 1 Heisenberg antiferromagnets on the spatially anisotropic square lattice has been scarcely explored. Here we report a study of the magnetism, specific heat, and thermal conductivity on Ni[SC(NH$_2$)$_2$]$_6$Br$_2$ (DHN) single crystals. Ni$^{2+}$ ions feature an $S$ = 1 rectangular lattice in the $bc$ plane, which can be viewed as an unfrustrated spatially anisotropic square lattice. A long-range antiferromagnetic order is developed at $T \rm_N =$ 2.23 K. Below $T\rm_N$, an upturn is observed in the $b$-axis magnetic susceptibility and the resultant minimum might be an indication for the $XY$ anisotropy in the ordered state. A gapped spin-wave dispersion is confirmed from the temperature dependence of the magnetic specific heat. Anisotropic temperature-field phase diagrams are mapped out and possible magnetic structures are proposed.

\end{abstract}

%75.50.-y Studies of specific magnetic materials
%66.70.-f Nonelectronic thermal conduction and heat-pulse propagation in solids
%75.47.-m Magnetotransport phenomena; materials for magnetotransport

\maketitle

\section{Introduction}

Two-dimensional (2D) square lattice Heisenberg antiferromagnets (HAFMs) have become a fertile field for condensed matter physics because of the prominent progress of high-temperature superconductors. In a square lattice, if the interaction along the diagonal $J_2$ becomes as significant as the one along the side of the square $J_1$, the competition between them frustrates the lattice and results in a rich phase diagram. In $S =$ 1/2 systems, three types of ordered ground states appear depending on the $J_2/J_1$ ratio \cite{BaCd}. When $|J_2/J_1| <$ 0.4, a N\'{e}el antiferromagnetic (AFM) state is formed for $J_1 >$ 0; in contrast, a ferromagnetic (FM) state is realized for $J_1 <$ 0. As $J_2$ becomes stronger, a columnar AFM state is stabilized for $|J_2/J_1| >$ 0.7 no matter what the sign of $J_1$ is. In the intermediate regions, the system is highly frustrated and novel quantum ground states are emerged, like spin liquid state for $J_1 >$ 0 and spin nematic state for $J_1 <$ 0. This $J_1-J_2$ model has been successfully applied to many $S =$ 1/2 compounds \cite{BaCd, Thesis, Extension, Pb2-1, Pb2-2, PbZn, LSVO-1, LSVO-2, LSVO-3, LSVO-4, PbVO3, VOMoO4, AMoOPO4Cl, MoOPO4}, while analogous study on $S$ = 1 systems has been rarely carried out. Quantum phase diagram of a spatially anisotropic $S$ = 1 square lattice has been established theoretically \cite{S=1_1,S=1_2,S=1_3}. The anisotropy factor is defined as $\alpha = J_{1x}/J_{1y}$, where $J_{1x}$ and $J_{1y}$ are the anisotropic nearest-neighbor couplings. When the frustration effect is negligible ($J_2$ = 0), the lattice can be viewed as decoupled Haldane chains for $\alpha$  = 0 and the ground state is gapped, while $\alpha$ = 1 corresponds to the isotropic square lattice with a N\'{e}el ground state. A quantum critical point separating the gapped and N\'{e}el states is proposed to locate at $\alpha\rm_c$ $\sim$ 0.05. With increasing $J_2$, a stripe order is introduced. As far as we know, few $S$ = 1 square lattice antiferromagnets have been discovered \cite{S=1_4,S=1_5}. In addition, these works mainly focused on the syntheses of new compounds, whereas the detailed magnetisms, phase diagrams, and magnetic transitions have not been investigated.

In this work, we grow single crystals of an $S =$ 1 rectangular lattice HAFM, Ni[SC(NH$_2$)$_2$]$_6$Br$_2$ (dibromo-hexakis thiourea-nickel(II), abbreviated as DHN), whose magnetic properties have not been reported yet. With careful structural analysis, the crystal structure is found to have highly spatial anisotropy. The magnetic exchange interaction is propagated through the -Ni-S$\cdots$S-Ni- pathway along the side of the rectangular lattice, which is absent along the diagonal direction. As a result, DHN can be considered as an unfrustrated spatially anisotropic $S$ = 1 square lattice antiferromagnet. A long-range AFM order is developed at $T\rm_N$ = 2.23 K determined by the magnetic susceptibility and specific heat measurements. Below $T\rm_N$, an upturn is observed in the $b$-axis magnetic susceptibility, and the resultant minimum is likely an indication for the $XY$ anisotropy in the ordered state. A gapped spin-wave dispersion is confirmed from the temperature dependence of the magnetic specific heat. Based on all experimental results, the anisotropic temperature-field phase diagrams are constructed and the possible magnetic structures are proposed.

\section{Experiments}

DHN single crystals were synthesized by evaporation method in a solution of water. 40 mmol thiourea was first dissolved in 30 ml deionized water, and then 10 mmol nickel bromide hydrate was added. Continuously stir the dark red solution until the solvents were dissolved completely. The solution was kept in water bath at 60 $^{\circ}$C for one hour, and then kept at 40 $^{\circ}$C to evaporate slowly. The crystals were harvested after about three weeks. The as-grown single crystals are dark green and opaque with irregular rectangle shape. The typical size of the crystal is (10--15) mm ${\times}$ (3--5) mm ${\times}$ (1.5--2) mm, as shown in the inset to Fig. 1(f).

Crystal structure was analyzed by four-circle x-ray diffraction at room temperature. Magnetic susceptibility and magnetization were measured at 1.8--300 K by using a SQUID-VSM (Quantum Design) and at 0.5--2 K by using a SQUID equipped with a $^3$He refrigerator (Quantum Design). Pulsed-field magnetization at 1.4 K was measured using an induction method at Wuhan National High Magnetic Field Center (China). Specific heat was measured by a relaxation method in the temperature range from 0.4 to 30 K using a PPMS (Quantum Design). Thermal conductivity was measured using a ``one heater, two thermometers" technique in a $^3$He refrigerator and a 14 T magnet at temperature regime of 0.3--8 K and using a Chromel-Constantan thermocouple in a $^4$He pulse-tube refrigerator in 0 T above 4 K \cite{Sun_DTN, Wang_HMO, Zhao_GFO, Song_CFO}. In the thermal conductivity measurements, the heat current is always applied along the $c$ axis, and the magnetic field is either along the $b$ or $c$ axis.

\section{Results}

\subsection{Crystal structure}

\begin{figure}
\includegraphics[clip,width=8.5cm]{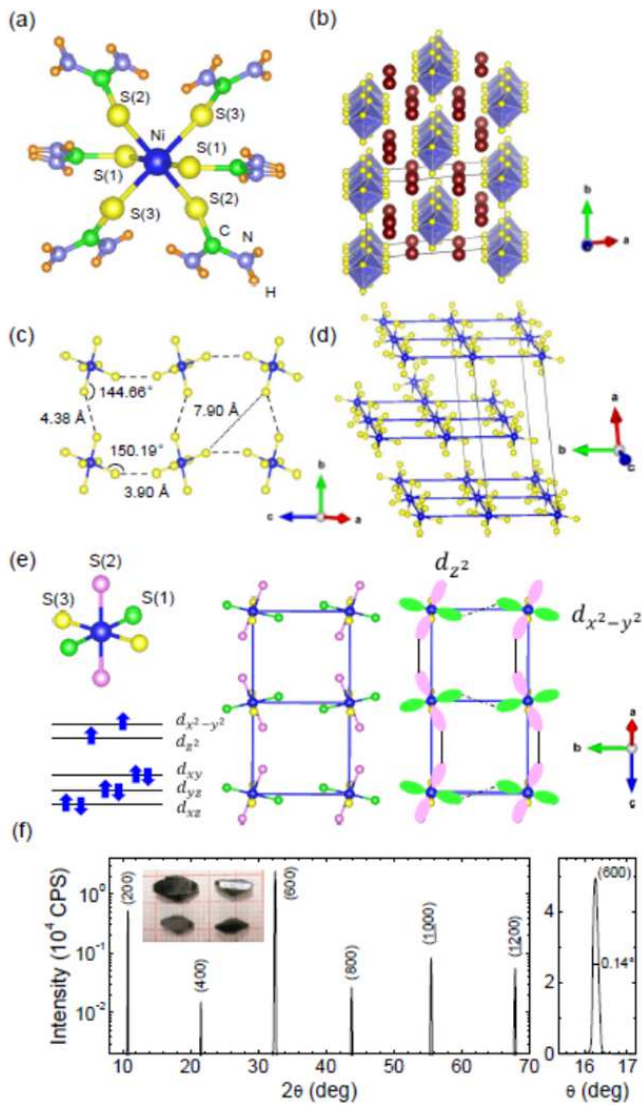}
\caption{(color online) (a) Local structure of DHN. (b) View of the crystal structure along the $c$ axis. The red balls are Br ions. The thiourea molecules are omitted for clarity. (c) Single layer in the $bc$ plane. Br and thiourea molecules are omitted for clarity. (d) Three dimensional crystal structure highlighting the $b/2$ displacement of the neighboring $bc$ layers. (e) Orbits of Ni ions projected in the $bc$ plane. Three particular S atoms are labeled by different colors. The pink ellipses denote $d_{z^2}$ orbit and the green ellipses denote the $x$ (or $y$) component of $d_{x^2-y^2}$ orbit. The dashed and dot-dashed lines are the exchange interaction through -Ni-S$\cdots$S-Ni- pathway. (f) X-ray diffraction on the ($h$00) facet and the rocking curve of the (600) diffraction. The width at the half maximum is 0.14$^\circ$. The inset is a photo of DHN single crystals.}
\end{figure}

DHN belongs to $M$(tu)$_6$$X_2$ family, in which $M$ is a transition metal, tu is shorted for thiourea SC(NH$_2$)$_2$, and $X$ is a halogen or NO$_3^-$ \cite{Ni[SC(NH2)2]6Br2}. DHN crystallizes in a monoclinic structure with $C2/c$ space group \cite{explaination}. The local structure is plotted in Fig. 1(a). Equivalent Ni$^{2+}$ ions are coordinated with six thiourea molecules forming Ni(tu)$_6$ octahedra, which are well separated by Br$^-$ ions as seen in Fig. 1(b). The Ni(tu)$_6$ octahedron has a weak trigonal distortion with three of the sulfur atoms moving toward each other. Each sulfur atom in thiourea molecules has a pair of electrons so as to contribute an $sp^2$ non-bonding orbit to form the Ni-S bond. The refined lattice parameters are list in Table I, and the atomic coordinates and the selected bond lengths and angles are given in Tables II and III in Appendix. All the S-C, C-N, and N-H bond lengths are close to the respective distances in the free thiourea molecule \cite{Ni[SC(NH2)2]6Br2}.

%\begin{table}[!ht]
%\caption{Lattice parameters of DHN.}
%\begin{ruledtabular}
%\begin{tabular}{llll}
%Formula & Ni[SC(NH$_2$)$_2$]$_6$Br$_2$ \\
%Formula weight & 675.28 \\
%Crystal system & Monoclinic \\
%Space group & $C2/c$ \\
%$a$ ({\AA}) & 22.898(2) \\
%$b$ ({\AA}) & 8.9087(3) \\
%$c$ ({\AA}) & 16.8040(17) \\
%$\alpha$ ($^{\circ}$) & 90 \\
%$\beta$ ($^{\circ}$) & 133.674(18) \\
%$\gamma$ ($^{\circ}$) & 90 \\
%$V$ ({\AA}$^3$) & 2479.3(8) \\
%$Z$ & 4 \\
%$\rho$ (g cm$^{-3}$) & 1.809 \\
%\end{tabular}
%\end{ruledtabular}
%\end{table}

\begin{table}[!ht]
\caption{Room temperature crystallographic data for DHN.}
\begin{threeparttable}
\begin{ruledtabular}
\begin{tabular}{llll}
Formula & Ni[SC(NH$_2$)$_2$]$_6$Br$_2$ \\
Formula weight & 675.28 \\
Wavelength ({\AA}) & 0.71073 \\
Crystal system & Monoclinic \\
Space group & $C2/c$ \\
$a$ ({\AA}) & 22.898(2) \\
$b$ ({\AA}) & 8.9087(3) \\
$c$ ({\AA}) & 16.8040(17) \\
$\alpha$ ($^{\circ}$) & 90 \\
$\beta$ ($^{\circ}$) & 133.674(18) \\
$\gamma$ ($^{\circ}$) & 90 \\
$V$ ({\AA}$^3$) & 2479.3(8) \\
$Z$ & 4 \\
$\rho$ (g cm$^{-3}$) & 1.809 \\
$\mu$ (mm$^{-1}$) & 4.531 \\
$R_1$, $wR_2$ ($F\rm_o >$ 4$\sigma$$F\rm_o$) & 0.0385, 0.0989 \\
$R_1$, $wR_2$ (all data) & 0.0453, 0.1057 \\
Goodness-of-Fit & 0.9581 \\
\end{tabular}
\end{ruledtabular}
\begin{tablenotes}
\footnotesize
\item[a] $R_1 = \Sigma\mid\mid F{\rm_o}\mid - \mid F{\rm_c}\mid\mid/\Sigma\mid F{\rm_o}\mid$, $wR_2 = [\Sigma w({F{\rm_o}}^2-{F{\rm_c}}^2)^2/\Sigma w({F{\rm_o}}^2)^2]^{1/2}$
\end{tablenotes}
\end{threeparttable}
\end{table}

In DHN, the Ni(tu)$_6$ octahedra are isolated from each other, and the neighbouring Ni$^{2+}$ ions interact with each other through the -Ni-S$\cdots$S-Ni- two-sulfur exchange pathway. It has been demonstrated that in most copper complexes with two-halide exchange the predominant coupling is dependent on the interhalide distance instead of the intercopper distance \cite{molecular_magnet_review}. Similarly, the S$\cdots$S distance is also an important parameter in DHN to judge the relative magnitude of the interaction along three directions. The shortest S$\cdots$S distance between the adjacent $bc$ planes is 5.77 {\AA}, much longer than those along the $b$ axis (4.38 {\AA}) and the $c$ axis (3.90 {\AA}), as shown Fig. 1(c). The crystal structure of DHN is thus constituted of Ni(tu)$_6$ layers stacked along the $a$ axis with a $b$/2 displacement alternately (Fig. 1(d)). The 2D magnetic correlation can also be understood from the orbital scenario. It is known that for a Ni$^{2+}$ ion in an octahedron crystal field the $d_{x^2-y^2}$ and $d_{z^2}$ orbits are magnetically active. Since the Ni-S(2) bond length is longer than Ni-S(1) and Ni-S(3), the local $z$ axis in a Ni(tu)$_6$ octahedron points to the S(2) atom. As illustrated in Fig. 1(e), the Ni-S(1) and Ni-S(2) bonds are almost lied within the $bc$ plane and point along the $c$ and $b$ directions, respectively. Consequently, the coupling $J_c$ along the $c$ axis is established through the $d_{z^2}$ orbit, while the $x$ (or $y$) component of $d_{x^2-y^2}$ orbit is responsible for the coupling $J_b$ along the $b$ axis. On the other hand, although the S$\cdots$S distance between adjacent layers is shorter than the diagonal one (7.90 {\AA}) in the $bc$ plane, the interlayer exchange interaction $J \rm_\perp$ should be weak when the exchange pathway is via the $y$ (or $x$) component of $d_{x^2-y^2}$ orbit due to the $b$/2 shift along $a$ axis. The nearly ferrodistortive order of Ni(tu)$_6$ octahedra in $bc$ plane results in a strong intralayer coupling, leaving a lack of possible orbital overlap along the perpendicular direction. This peculiar stacked arrangement of the crystal structure makes DHN favorable to form a spatially anisotropic 2D lattice, which is confirmed by the high-temperature magnetic susceptibility as discussed below.

The crystal structure of DHN is reminiscent of NiCl$_2$-4SC(NH$_2$)$_2$ (DTN), a famous compound exhibiting a magnon Bose-Einstein condensation \cite{Sun_DTN, DTN-2}. The structure-property relation between these two compounds is worthy of addressing. DTN is a member of $M$(tu)$_4$$X_2$, which has the similar chemical composition to DHN except for the replacement of Cl by Br. In DTN, equivalent Ni$^{2+}$ ion is surrounded by two Cl$^-$ and four thiourea molecules forming a NiCl$_2$(tu)$_4$ octahedron. All NiCl$_2$(tu)$_4$ octahedra are isolated, and the superexchange propagates via -Ni-Cl$\cdots$Cl-Ni- and -Ni-S$\cdots$S-Ni- pathways. Neutron scattering measurement detected that the Ni spins are strongly coupled along the -Ni-Cl$\cdots$Cl-Ni- pathway, making DTN a weakly coupled spin-chain system \cite{DTN-2}. In DHN, however, Ni$^{2+}$ ions are coordinated with six thiourea molecules, and then the system behaves as a 2D HAFM.

In the $bc$ plane, DHN features a rectangular lattice of Ni$^{2+}$ ions due to the inequivalent nearest-neighboring (NN) interactions. The frustration as well as the coupling $J_2$ along the diagonal direction should be very weak for two reasons. First, since the S$\cdots$S distance along the diagonal is 7.90 {\AA}, it is likely too long for an efficient orbital overlap between the thiourea groups. Second, no bridging ligands are present to establish $J_2$. In the frustrated vanadium and molybdenum oxides, both $J_1$ and $J_2$ are formed through the corner-shared PO$_4$ \cite{BaCd,Extension,PbZn}, (Si, Ge)O$_4$ \cite{LSVO-1,LSVO-2,LSVO-3}, or MoO$_4$ \cite{AMoOPO4Cl,MoOPO4} tetrahedra, resulting in a comparable NN and next-nearest-neighbor (NNN) interactions. However, such bridging ligands are missing in DHN. Similar situation is found in some weakly frustrated pyrazine-based copper complexes, in which the orientation of the magnetic $d_{x^2-y^2}$ orbit occupying the basal plane gives a direct overlap with the ligand orbit of the pyrazine unit along the side of the square \cite{pz-1, pz-2, pz-3, pz-4, crossover-2, crossover-3}. The absence of the orbital overlap along the diagonal suggests that DHN is free from the frustration effect. It is known that the S$\cdots$S non-bonding contact is sensitive to the S$\cdots$S distance and $M$-S$\cdots$S bond angle \cite{CCR}. The larger the bond angle and the shorter the distance are, the stronger the AFM exchange interaction is. From Fig. 1(c) one can expect that the inequivalent $J_b$ and $J_c$ are both AFM with $J_c > J_b$. As a result, DHN can be regarded as an unfrustrated spatially anisotropic $S$ = 1 square lattice antiferromagnet.

The inset to Fig. 1(f) is a photograph of several single crystals, which exhibit some large natural surfaces. The largest facet of the single crystal is the $bc$ plane, which was confirmed by the x-ray diffraction as seen in Fig. 1(f). The good crystallinity was also verified by the narrow rocking curve. More diffractions on different natural surfaces further confirmed that the longest dimension of single crystal is the $c$ axis.

\subsection{Magnetic properties}

Figure 2(a) shows the temperature dependencies of the magnetic susceptibility $\chi(T)$ measured in an applied field of 1 T parallel to the $b$ axis, $c$ axis, and normal to the $bc$ plane (labeled as $\chi_\bot$). Above 3 K, both $\chi_b$ and $\chi_c$ are larger than $\chi_\bot$, which suggests a $g$-factor anisotropy resulting from the 2D magnetic correlation within the $bc$ plane as proposed from the forementioned structural analysis \cite{crossover-1, anisotropic_g-1, anisotropic_g-2}. $\chi(T)$ increases with decreasing temperature in a Curie-Weiss manner $\chi = C/(T - \theta \rm_{CW})$. The fit for $\chi_c$ gives $C =$ 1.45 emu K/mol and $\theta \rm_{CW} =$ -7.17(2) K. The negative $\theta \rm_{CW}$ indicates a dominant AFM interaction, and the effective moment is deduced to be 3.41 $\mu \rm_B$/f.u. A broad maximum is present around 4 K, which is a hallmark for the low-dimensional magnets and signals the buildup of 2D AFM correlation.

\begin{figure}
\includegraphics[clip,width=8.0cm]{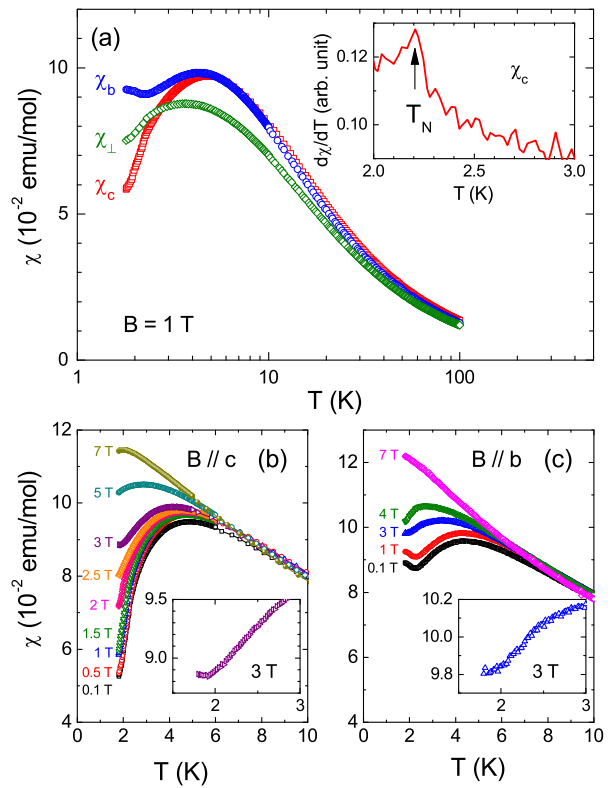}
\caption{(color online) (a) Magnetic susceptibility of DHN crystals measured in 1 T field along different directions. The inset is a differential of $\chi_c$. The onset temperature of the AFM transition $T\rm_N$ is highlighted by the arrow. (b,c) Field dependencies of $\chi_c$ and $\chi_b$ at low temperatures from 0.1 to 7 T. The insets are the zoom-in of 3 T data at $T <$ 3 K.}
\end{figure}

As further cooling, a weak slope change is observed in both $\chi_c$ and $\chi_\bot$ at 2.2 K, which is clearly seen from the differential of $\chi_c$ as shown in the inset to Fig. 2(a). As confirmed by the $\lambda$ peak in the specific heat (see below), this slope change is associated with the long-range AFM order, which is driven by the inevitable interlayer interaction. The weakness of $T\rm_N$ observed from $\chi(T)$ is usually found in low-dimensional antiferromagnets \cite{BaCd, Pb2-2, Thesis}. The transition temperature is suppressed rapidly with increasing the magnetic field. In 1.5 T field, $T \rm_N$ is decreased to 2.11 K and it is almost disappeared above 2.5 T.

For 2D HAFMs, the effective exchange interaction can be roughly estimated by using the following high temperature series expansion (HTSE) for arbitrary $S$ in any lattice \cite{HTSE},
\begin{equation}
\chi = \frac{Ng^2\mu\rm_B^2}{3k\rm_B\textit{T}}S(S+1)(1 + \sum a_n x^n)^{-1},
\end{equation}
here $S =$ 1, $x = J/k\rm_B$$T$, $k\rm_B$ is the Boltzmann constant, $N$ is the Avogadro's constant, $g$ is the Land\'{e} factor, $\mu\rm_B$ is the Bohr magneton, and $a_n$ is the free parameter. This relation has been successfully applied to $A$MoOPO$_4$Cl, and the fitting is as good as the $J_1-J_2$ model \cite{AMoOPO4Cl}. In DHN, $\chi_c$ can be well reproduced above 15 K by this formula and yields $J =$ 45(6) K. However, the obtained $J$ is unreasonable since it is too large for an organic complex through the two-sulfur exchange pathway \cite{pz-1}, which might be due to the exclusion of the single-ion anisotropy of Ni$^{2+}$ ions in the model.

Below $T \rm_N$, $\chi_c$ and $\chi_\bot$ are decreased continuously while an upturn is emerged in $\chi_b$. The presence of the minimum is likely an indication for the $XY$ anisotropy as observed in several $S$ = 1/2 square lattice compounds \cite{crossover-1,AMoOPO4Cl,Pb2-2}. It was proposed theoretically \cite{XY_behavior-1, XY_behavior-2} that as the $XY$ anisotropy becomes significant, a large amount of spins anti-align in the plane. As a consequence, the in-plane component decreases more quickly, while the out-of-plane component slows down its decrease with weak ferromagnetic component which results in a minimum at lower temperatures. Since the minimum is displayed in $\chi_b$, the $XY$ plane in DHN is found to be the $ac$ plane. For clarity, in the following discussion we only show the properties for $B \parallel c$ to represent the $XY$-plane behavior.

\begin{figure}
\includegraphics[clip,width=8.5cm]{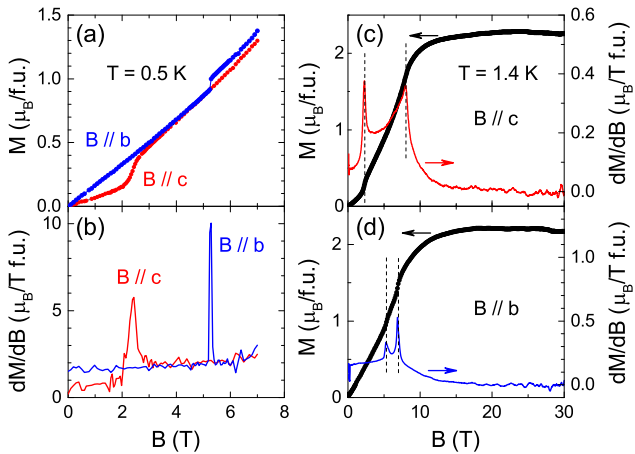}
\caption{(color online) Anisotropic magnetization and differentials of DHN crystals in static field at 0.5 K (a,b) and in pulsed field at 1.4 K (c,d).}
\end{figure}

The anisotropic magnetization is displayed in Fig. 3. At low fields, $M_c$ ($B \parallel c$) is much smaller than $M_b$ ($B \parallel b$). As the field is increased, $M_c$ exhibits a quick enhancement around 2.5 T. The extrapolation from the high fields goes across the origin, suggesting a spin-flop transition without a generated net moment. It is known that when the magnetic field is applied along the easy axis, a spin-flop transition can occur above a critical field $B \rm_{SF}$ with all spins re-orientate to the perpendicular direction. After that, spins will gradually point to the field direction and reach the saturation at $B \rm_c$. In the molecular field model, both the critical fields are related to the exchange field $B \rm_E$ and the anisotropic field $B \rm_A$ as below \cite{anisotropy},
\begin{eqnarray}
B \rm_{SF} &=& (2\textit{B}\rm_E \textit{B}\rm_A - {\textit{B}\rm_A}^2)^{1/2}\\
B \rm_c &=& 2\textit{B}\rm_E - \textit{B}\rm_A.
\end{eqnarray}
To obtain $B \rm_E$ and $B \rm_A$, pulsed-field magnetization is performed to determine $B \rm_{SF}$ and $B \rm_c$. As seen in Fig. 3(c), $B \rm_{SF} =$ 2.25 T and $B \rm_c =$ 8 T are determined for $B \parallel c$. According to the model, one can get $B \rm_E =$ 4.32 T and $B \rm_A =$ 0.63 T. The ratio of $B \rm_A$/$B \rm_E$ is about 0.15, which indicates a strong anisotropy in the $XY$ plane. In most layered compounds, this ratio is usually less than 1\% \cite{Layered_compound_review}. It is in agreement with the magnetic susceptibility, in which the drop of $\chi_a$ is less significant than $\chi_c$, as shown in Fig. 2(a). As discussed above, spins are confined in the $ac$ plane due to the $XY$ anisotropy. When the field is increased to 3 T, just above the spin-flop transition, the minimum in $\chi_b$ disappears, and meanwhile $\chi_c$ shows a minimum at 1.9 K, as shown in the insets to Figs. 2(b) and 2(c). This suggests that the $XY$ plane changes from $ac$ to the plane normal to the $c$ axis, and $c$ axis becomes the hard axis due to the presence of the minimum under the influence of the magnetic field.

In contrast to $M_c$, $M_b$ shows a step-like increase at 5 T, above which the magnetization increases linearly up to 7 T, suggesting a strong AFM correlation. The extrapolation to zero field is positive, revealing a weak ferromagnetic component induced in high fields. Considering the spin-flop transition in 2D systems is commonly broad due to the domain wall \cite{Layered_compound_review}, the observed sharp enhancement in $M_b$ is more likely a spin re-orientation to an AFM configuration with a weak ferromagnetic (WFM) component. The corresponding critical fields from the pulsed-field magnetization are $B \rm_{m} =$ 5.26 T and $B \rm_c =$ 6.91 T.

\subsection{Specific heat}

\begin{figure}
\includegraphics[clip,width=6.5cm]{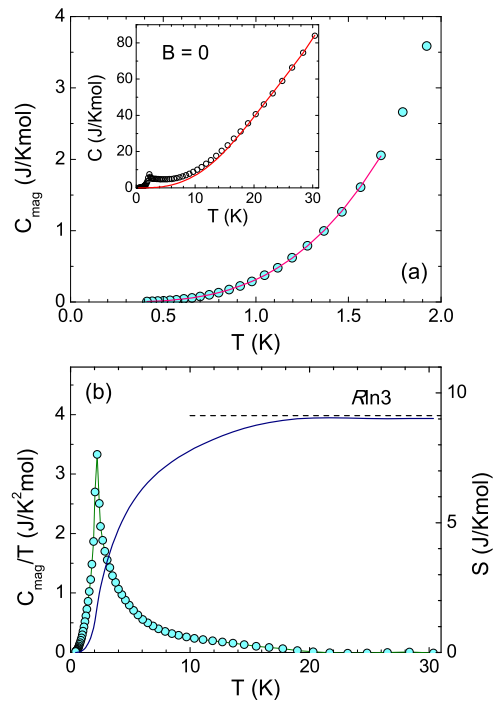}
\caption{(color online) (a) Magnetic specific heat of DHN crystal in zero field. Inset is the total specific heat measured in 0 T. The solid lines are fitting results as discussed in the text. (b) The temperature dependencies of $C \rm_{mag}$/$T$ and the entropy $S$. The horizontal line is the theoretical entropy recovery for an $S =$ 1 system.}
\end{figure}

The inset to Fig. 4(a) shows the total specific heat $C$ measured in zero field between 0.4 and 30 K. The $\lambda$-shaped anomaly observed at $T \rm_N =$ 2.23 K is associated with the AFM order as referred in the magnetic susceptibility. The lattice specific heat $C\rm_L$ can be estimated by using the low-frequency expansion of the Debye function
\begin{equation}
C\rm_L = \beta_3\textit{T}^3 + \beta_5\textit{T}^5 + \beta_7\textit{T}^7,
\end{equation}
where $\beta_3$, $\beta$$_5$, and $\beta$$_7$ are the temperature-independent parameters \cite{Cp}. As can be seen in the inset to Fig. 4(a), the experimental data above 20 K can be fitted by this formula quite well with $\beta_3 =$ 7.6(1)$\times$10$^{-3}$ J/K$^4$mol, $\beta_5 =$ -8.3(4)$\times$10$^{-6}$ J/K$^6$mol, and $\beta_7 =$ 3.5(2)$\times$10$^{-9}$ J/K$^8$mol. The Debye temperature deduced from $\beta_3$ is 235 K using $\theta \rm_D ^3 =$ 12$\pi^4$$Rs$/5$\beta$, where $R$ is the gas constant and $s =$ 51 is the number of atoms per molecule. The magnetic specific heat $C \rm_{mag}$ can then be extracted by subtracting the lattice contribution from the experimental data, as shown in the main panel of Fig. 4(a). The entropy is completely recovered to the expected value of $R$ln3 for $S =$ 1 systems. The magnitude at $T \rm_N$ is 1.8 J/Kmol and is about 20\% of the saturation, demonstrating that the residual entropy is almost consumed above $T \rm_N$ due to the 2D short-range correlation.

Below $T \rm_N$, $C \rm_{mag}$ deviates from the gapless $T^3$ dependence for 3D AFM order. When there is an anisotropic gap in the spin wave dispersion, an exponential term is appeared in the magnetic specific heat \cite{Cp}
\begin{equation}
C \rm_{mag} = \textit{a}\textit{T}^3exp(-\Delta/\textit{T}).
\end{equation}
As shown in Fig. 4(a), $C \rm_{mag}$ below 1.7 K can be described well using this relationship, and the fit gives $\Delta =$ 0.81(1) K. This energy gap should be originated from the presence of the $XY$ anisotropy as observed in the magnetic susceptibility.

\begin{figure}
\includegraphics[clip,width=8.0cm]{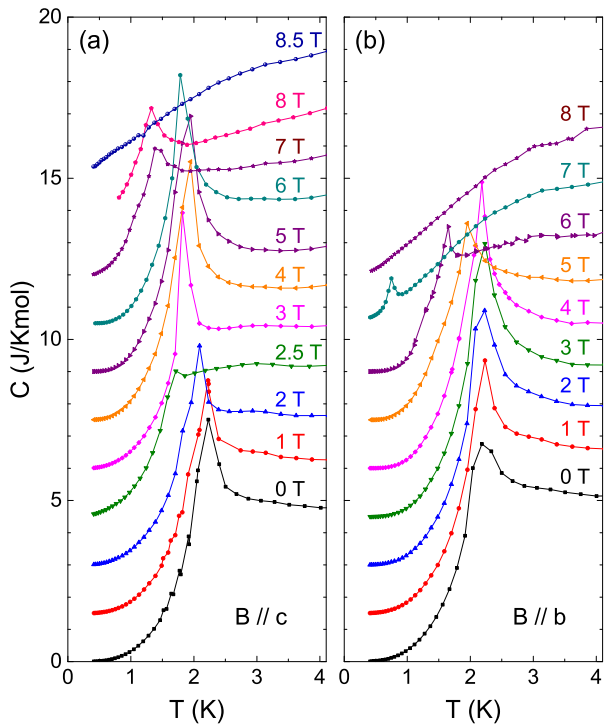}
\caption{(color online) Specific heat of DHN crystals in magnetic fields along the $b$ and $c$ axis, respectively. For clarity, the in-field specific heat is shifted upword by 1.5 J/Kmol in sequence.}
\end{figure}

When applying magnetic field along different directions, the specific heat changes in different ways for $B \parallel c$ and $B \parallel b$, as shown in Fig. 5. For $B \parallel c$, below 2 T the $\lambda$ peak moves slightly toward low temperatures, and the magnitude is decreased significantly in 2.5 T. With increasing field, another sharp peak appears at 1.8 K in 3 T, and the peak position is enhanced to 1.9 K in 5 T. After that, the peak is suppressed again and disappears completely at 8.5 T. It is clear that the low-field peak is related to the transition from the paramagnetic state to the N\'{e}el state, and the high-field peak is related to the transition to the spin-flop state. When the field is applied along the $b$ axis, the situation is much simpler: $T \rm_N$ is first slightly increased up to 3 T and then decreased at 4 T; when entering into the WFM state, the $\lambda$ peak is strongly suppressed and moved rapidly toward low temperatures; above 8 T, the magnetic order disappears completely.

\subsection{Thermal conductivity}

\begin{figure}
\includegraphics[clip,width=6.5cm]{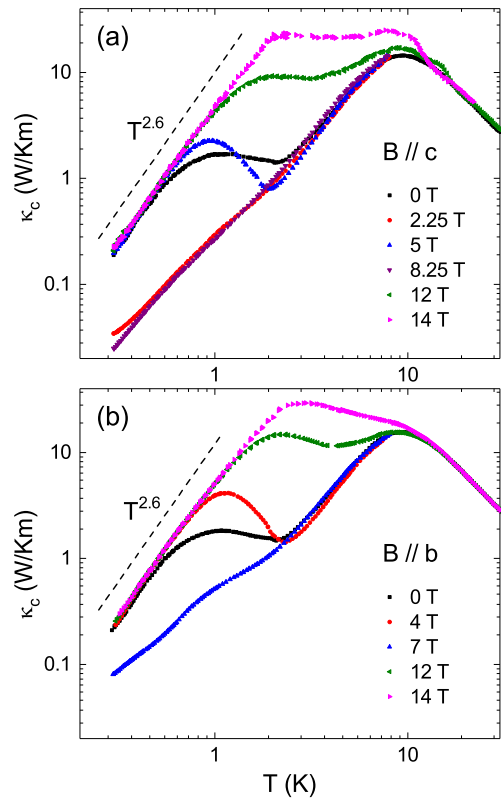}
\caption{(color online) Temperature dependence of the thermal conductivity of DHN crystal in different magnetic fields applied along the $b$ or $c$ axis.}
\end{figure}

Figure 6 shows the temperature dependencies of thermal conductivity in zero field and with  magnetic fields applied along the $b$ or $c$ axis. In zero filed, the $\kappa$ shows a double-peak behavior with a minimum at about 2 K, which is originated from the strong phonon scattering across the AFM transition \cite{Wang_HMO, Song_CFO}. The magnitude of the left shoulder is strongly field dependent. Especially near the transition fields in $M(B)$, the magnitude is nearly decreased by one order at the lowest temperature for $B \parallel c$ and the double-peak feature is smeared out. When the field is higher than the polarization field, the $\kappa$ starts to recover in a wide temperature range due to the weakened phonon scattering. In 14 T, the scattering is strongly suppressed and the double-peak feature becomes much weaker. It can be expected that a single large phonon peak can be recovered in high enough fields \cite{Song_CFO}. The disappearance of the phonon scattering by magnon can also be evidenced by the temperature dependence of the subKelvin $\kappa$ in 14 T, which shows an approximate $T^{2.6}$ dependence and indicates that the phonon boundary scattering is nearly approached \cite{Berman}.

\begin{figure}
\includegraphics[clip,width=8.5cm]{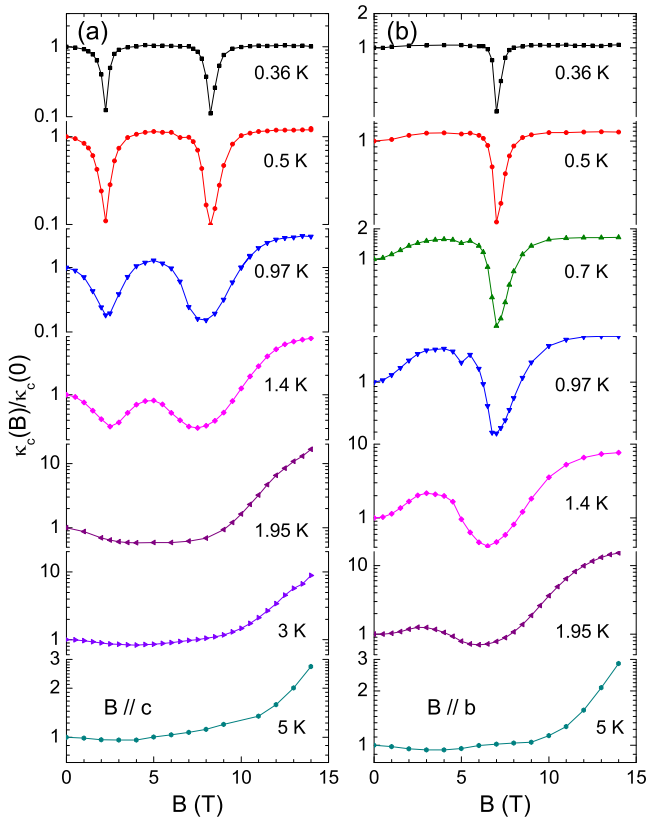}
\caption{(color online) Magnetic-field dependence of the thermal conductivity of DHN crystal at low temperatures.}
\end{figure}

Figure 7 shows the detailed magnetic-field dependencies of thermal conductivity at low temperatures. It is clearly seen that the coupling between phonons and magnetic excitations is so strong that $\kappa$ exhibits significant field dependence, that is, the magnitude can be suppressed down to as low as $\sim$ 10\% and recovered up to 15 times in 14 T. There are two sharp dips observed for $B \parallel c$, which are related to the spin-flop and spin polarization transitions, respectively, as demonstrated from the magnetization and specific heat. It is known that increasing field can overcome the anisotropy and induce a spin reorientation with a closed gap at these critical fields. Consequently, the phonon scattering by the populated magnon excitations is the strongest, resulting in a dip-like feature in the $\kappa(B)$ isotherms \cite{Wang_HMO, Zhao_GFO}. The dip-like feature is nearly field independent below 0.7 K and becomes broader at higher temperatures. Above 2 K, this feature completely vanishes and the $\kappa$ is monotonically increased with increasing field, demonstrating a significant suppression of the magnetic excitations. Note that the strongest enhancement of $\kappa$ is occurred at 1.95 K, in which $\kappa$ in 14 T is almost 15 times of the zero-field value. Similar dip-like feature is also observed for $B \parallel b$, but lower-field dip related to the spin re-orientation is rather weak at 0.36 K, which becomes distinct only at 0.97 K.

\section{Discussion}

\begin{figure}
\includegraphics[clip,width=6.5cm]{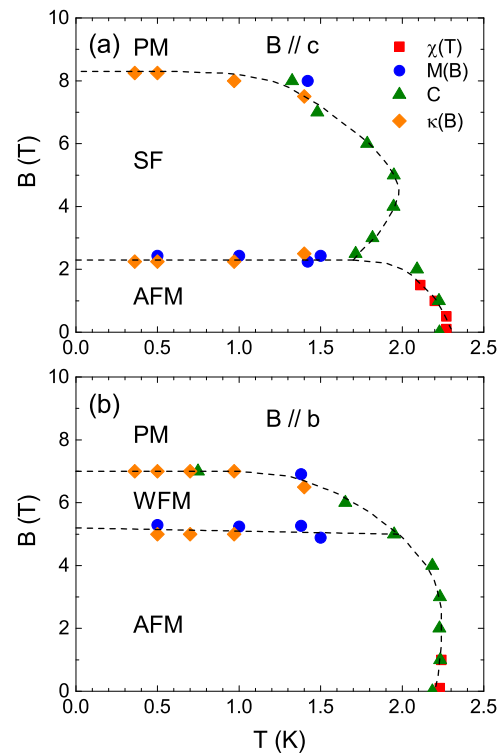}
\caption{(color online) $B-T$ phase diagrams of DHN. AFM, SF, WFM, and PM represents antiferromagnetic, spin flop, weak ferromagnetic, and paramagnetic state, respectively.}
\end{figure}

Based on the above experimental results, the $B-T$ phase diagrams for $B \parallel c$ and $B \parallel b$ are mapped out respectively, as shown in Fig. 8. As predicted in the anisotropic $XY$ square lattices \cite{Layered_compound_review, Phase_diagram-1, Phase_diagram-2}, a bicritical point, where the N\'{e}el state, the spin-flop state, and the paramagnetic states meet, is observed when the field is applied within the $XY$ plane. On the other hand, a transition to WFM state is unexpectedly occurred for $B \parallel b$.

The nonmonotonic field dependence of the ordering temperature has been reported in many quasi-2D magnets \cite{crossover-2, crossover-3, Phase_diagram-3, Phase_diagram-4}, and it was proposed to originate from the 2D quantum fluctuations even though the interlayer interaction is relatively strong \cite{Phase_diagram-3, XY_behavior-3}. When the field is weak, the quantum fluctuations of the out-of-plane component can be suppressed with increasing field. As a result, the $XY$ anisotropy is enhanced accompanied with the rise of $T \rm_N$. When the field is further increased, the spin-canting effect prevails. Then, $T \rm_N$ starts to decrease and eventually vanishes when the spins are fully polarized \cite{crossover-3}.

\begin{figure}
\includegraphics[clip,width=8.5cm]{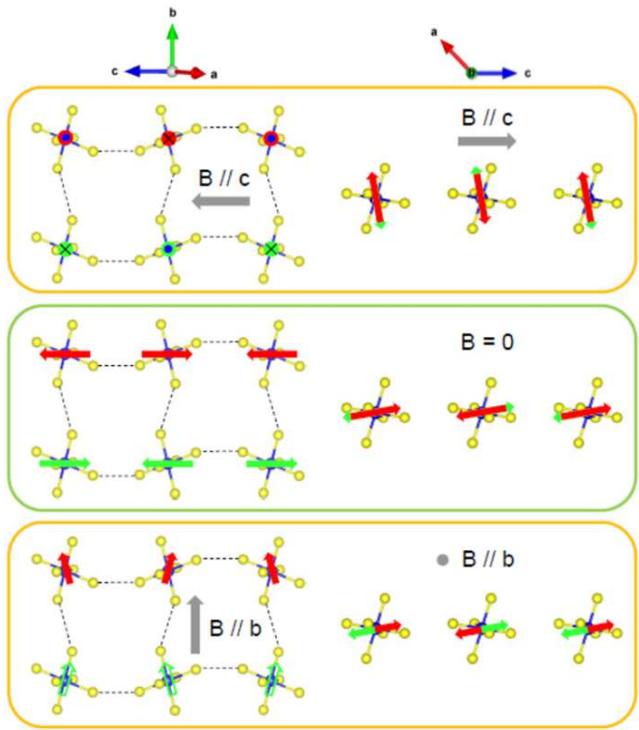}
\caption{(color online) Schematic magnetic structures projected in the $bc$ (left) and $ac$ (right) planes. The length of the arrow denotes the magnitude of the spin moment. Red and green arrows are the spins locating at different rows. The dot and cross in the upper panel denote the spin direction pointing ``into" and ``out" of the $bc$ plane. The small solid (open) arrows in the lower panel denote the spins having component pointing out of (into) the plane.}
\end{figure}

The possible spin structures projected in the $bc$ and $ac$ planes with different magnetic field directions are proposed and illustrated in Fig. 9. In zero field, the spins are aligned antiferromagnetically in the $ac$ plane and dominantly point to the $c$ direction. When a magnetic field is applied along the $c$ axis, a spin-flop transition is induced, and the spins are rotated perpendicular to the $c$ direction in the $ac$ plane. On the other hand, a spin re-orientation with a WFM component along the $b$ axis is induced with $B \parallel b$. The AFM correlation is still persisted considering the linear increase after the step-like enhancement in the magnetization.

\section{Summary}

In this work, we performed detailed structural and magnetism characterizations on DHN single crystals. In this compound, Ni$^{2+}$ ions feature an $S$ = 1 rectangular lattice, which can be regarded as a frustration-free spatially anisotropic square lattice. Due to the inevitable interlayer interaction, a long-range AFM order occurs at $T \rm_N$ = 2.23 K. Below $T \rm_N$, a minimum is observed in $\chi_b$ which is likely an indication for the $XY$ anisotropy in the AFM state. The temperature dependence of the magnetic specific heat evidences a gapped spin wave dispersion, which is also likely originated from the $XY$ anisotropy. DHN is therefore of great interest since there are few $S =$ 1 examples on the unfrustrated square lattice and the $XY$ anisotropy has been rarely studied in $S =$ 1 2D lattices. Theoretical model and numerical calculation on $S$ = 1 square lattice are in urgent need to explain the novel properties of DHN in near future.

\begin{acknowledgements}

We thank Y. Kamiya for the helpful discussion. This work was supported by the National Natural Science Foundation of China (Grant Nos. 11574286, 11874336, U1832209, and U1532147), the National Basic Research Program of China (Grant Nos. 2015CB921201 and 2016YFA0300103), and the Innovative Program of Development Foundation of Hefei Center for Physical Science and Technology. Z.Y.Z. acknowledges the supports from the National Natural Science Foundation of China (Grant Nos. 51702320 and U1832166). H.D.Z. thanks for the support from the Ministry of Science and Technology of China with grant number 2016YFA0300500 and from NSF-DMR with grant number NSF-DMR-1350002.

\end{acknowledgements}

\section*{Appendix}

\begin{table*}[!ht]
\caption{Atomic coordinates and equivalent isotropic displacement parameters of DHN.}
\begin{ruledtabular}
\begin{tabular} {lllllllllll}
Atom & $x$ & $y$ & $z$ & $U^2$ & & Atom & $x$ & $y$ & $z$ & $U^2$\\
\cline{1-11} \\[0.1ex]
Ni & 0.50000 & 0.5000 & 0.5000 & 0.0243 & & H(1) & 0.3721 & 0.2621 & 0.4346 & 0.0561 \\
Br & 0.33105(1) & 0.97469(3) & 0.54756(2) & 0.0415 & & H(2) & 0.2947 & 0.1751 & 0.3453 & 0.0565 \\
S(1) & 0.35680(3) & 0.50987(6) & 0.32143(4) & 0.0312 & & H(3) & 0.2221 & 0.3844 & 0.1385 & 0.0772 \\
S(2) & 0.52299(3) & 0.60490(6) & 0.38543(4) & 0.0315 & & H(4) & 0.1981 & 0.2709 & 0.1765 & 0.0776 \\
S(3) & 0.47658(3) & 0.76743(5) & 0.51310(4) & 0.0301 & & H(5) & 0.6966 & 0.6129 & 0.4328 & 0.0546 \\
C(1) & 0.30153(12) & 0.3528(3) & 0.29092(17) & 0.0356 & & H(6) & 0.6281 & 0.7282 & 0.3819 & 0.0561 \\
C(2) & 0.61163(12) & 0.5487(2) & 0.42503(16) & 0.0300 & & H(7) & 0.6183 & 0.3549 & 0.4778 & 0.0513 \\
C(3) & 0.53276(12) & 0.8319(2) & 0.64486(16) & 0.0302 & & H(8) & 0.6935 & 0.3985 & 0.5043 & 0.0521 \\
N(1) & 0.32579(13) & 0.2507(3) & 0.36488(18) & 0.0479 & & H(9) & 0.5797 & 1.0076 & 0.7331 & 0.0612 \\
N(2) & 0.23153(14) & 0.3324(4) & 0.18919(18) & 0.0682 & & H(10) & 0.5326 & 1.0359 & 0.6139 & 0.0617 \\
N(3) & 0.65025(13) & 0.6421(3) & 0.4138(2) & 0.0456 & & H(11) & 0.5846 & 0.7773 & 0.7891 & 0.0510 \\
N(4) & 0.64189(12) & 0.4130(2) & 0.46416(19) & 0.0425 & & H(12) & 0.5547 & 0.6451 & 0.7168 & 0.0504 \\
N(5) & 0.54717(17) & 0.9765(2) & 0.66513(19) & 0.0520 \\
N(6) & 0.56079(14) & 0.7405(2) & 0.72621(16) & 0.0430 \\
\end{tabular}
\end{ruledtabular}
\end{table*}

\begin{table*}[!ht]
\caption{Selected bond lengths ({\AA}) and bond angles (degree) of DHN.}
\begin{ruledtabular}
\begin{tabular} {llllll}
Ni-S(1) & 2.4849(5) & C(2)-N(3) & 1.319(4) & N(3)-H(5) & 0.910(3) \\
Ni-S(2) & 2.5013(4) & C(2)-N(4) & 1.321(3) & N(3)-H(6) & 0.870(3) \\
Ni-S(3) & 2.4850(5) & C(3)-N(5) & 1.315(3) & N(4)-H(7) & 0.885(3) \\
S(1)-C(1) & 1.711(3) & C(3)-N(6) & 1.316(4) & N(4)-H(8) & 0.8760(13) \\
S(2)-C(2) & 1.717(3) & N(1)-H(1) & 0.890(3) & N(5)-H(9) & 0.872(4) \\
S(3)-C(3) & 1.718(3) & N(1)-H(2) & 0.863(3) & N(5)-H(10) & 0.858(3) \\
C(1)-N(1) & 1.314(4) & N(2)-H(3) & 0.856(3) & N(6)-H(11) & 0.852(3) \\
C(1)-N(2) & 1.319(4) & N(2)-H(4) & 0.840(4) & N(6)-H(12) & 0.8581(18) \\
\\
S(1)-Ni-S(2) & 82.00(4) & S(1)-Ni-S(3) & 83.54(5) & Ni-S(1)$\cdots$S(1) & 124.21(4) \\
S(1)-Ni-S(2) & 98.00(4) & S(2)-Ni-S(3) & 96.72(4) & Ni-S(2)$\cdots$S(2) & 150.19(5) \\
S(1)-Ni-S(3) & 96.46(5) & S(2)-Ni-S(3) & 83.28(4) & Ni-S(3)$\cdots$S(3) & 144.66(4) \\
\end{tabular}
\end{ruledtabular}
\end{table*}


\begin{thebibliography}{}
\bibitem{explaination}
A different space group $I2/c$ has been reported in Ref. [\onlinecite{Ni[SC(NH2)2]6Br2}]. We will not discuss the difference between these two structures. The crystallographic axes used in this work are determined using $C2/c$ space group.
\end{thebibliography}
\end{document}